\documentclass[journal,final]{IEEEtran}
\usepackage{epsfig,rotating,setspace,latexsym,amsmath,epsf,amssymb,bm}
\usepackage{color}

\usepackage{amsmath}
\usepackage{rotating,latexsym,amsmath,amssymb,bm}
\usepackage{cite}
\usepackage{amsthm}
\usepackage[caption=false,font=footnotesize]{subfig}
\usepackage{graphicx}

\def\nb0{{\mathbf{0}}}
\def\nb1{{\mathbf{1}}}

\def\ncalF{{\mathcal{F}}}

\def\ncalM{{\mathcal{M}}}
\def\ncalN{{\mathcal{N}}}
\def\ncalO{{\mathcal{O}}}
\def\ncalP{{\mathcal{P}}}
\def\ncalQ{{\mathcal{Q}}}

\def\ncalS{{\mathcal{S}}}

\def\ncalW{{\mathcal{W}}}

\def\ncalY{{\mathcal{Y}}}
\def\ncalZ{{\mathcal{Z}}}

\def\nbbR{{\mathbb{R}}}

\newtheorem{ndef}{Definition}

\newtheorem{theorem}{Theorem}

\def\E{\mathbb{E}}

\def\V{\operatorname{Var}}

\def\sinr{\mathtt{SINR}}			

\usepackage{bbm}
\usepackage{url}

\usepackage{wrapfig}
\usepackage{dsfont}
\usepackage{wasysym}
\usepackage{hyperref}
\usepackage{balance}
\usepackage[ruled,vlined]{algorithm2e}

\theoremstyle{plain}

\theoremstyle{definition}
\newtheorem{assum}{Assumption}

\usepackage{thmtools}

\declaretheoremstyle[
  spaceabove=\topsep, spacebelow=\topsep,
  headfont=\normalfont\bfseries,
  notefont=\mdseries, notebraces={(}{)},
  bodyfont=\normalfont,
  postheadspace=1em,
  qed=\qedsymbol
]{mythmstyle}

\declaretheoremstyle[
  spaceabove=\topsep, spacebelow=\topsep,
  headfont=\normalfont\bfseries,
  notefont=\mdseries, notebraces={(}{)},
  bodyfont=\normalfont,
  postheadspace=1em,
  qed=$\diamond$
]{mythmstyle}
\declaretheorem[style=mythmstyle]{remark}

\DeclareMathOperator{\lA}{\langle}
\DeclareMathOperator{\rA}{\rangle}

\begin{document}

\allowdisplaybreaks

\sloppy

\title{Online Bayesian Meta-Learning for \\ Cognitive Tracking Radar}

\author{Charles E. Thornton$^{*}$, R. Michael Buehrer$^{*}$, and Anthony F. Martone$^{\dagger}$
\thanks{$^{*}$Department of ECE, Virginia Tech, Blacksburg, VA, USA, 24061 (\emph{emails: \texttt{\{thorntonc,buehrer\}@vt.edu}}). $^{\dagger}$US Army Research Laboratory, Adelphi, MD, USA, 20783 (\emph{email: \texttt{anthony.f.martone.civ@army.mil}}). Portions of this work were presented at the 2022 IEEE Radar Conference \cite{Thornton2022}. The authors gratefully acknowledge support from the US Army Research Office (ARO).}}

\maketitle
\thispagestyle{plain}
\pagestyle{plain}
\vspace{-1cm}
\begin{abstract}
A key component of cognitive radar is the ability to \emph{generalize}, or achieve consistent performance across a range of sensing environments, since aspects of the physical scene may vary over time. This presents a challenge for learning-based waveform selection approaches, since transmission policies which are effective in one scene may be highly suboptimal in another. We address this problem by strategically \emph{biasing} a learning algorithm by exploiting high-level structure across tracking instances, referred to as \emph{meta-learning}. In this work, we develop an online meta-learning approach for waveform-agile tracking. This approach uses information gained from previous target tracks to speed up and enhance learning in new tracking instances. This results in sample-efficient learning across a class of finite state target channels by exploiting inherent similarity across tracking scenes, attributed to common physical elements such as target type or clutter statistics. We formulate the online waveform selection problem within the framework of Bayesian learning, and provide prior-dependent performance bounds for the meta-learning problem using Probability Approximately Correct (PAC)-Bayes theory. We present a computationally feasible meta-posterior sampling algorithm and study the performance in a simulation study consisting of diverse scenes. Finally, we examine the potential performance benefits and practical challenges associated with online meta-learning for waveform-agile tracking.
\end{abstract}

\begin{IEEEkeywords}
meta-learning, radar signal processing, target tracking, cognitive radar, statistical learning theory
\end{IEEEkeywords}

\section{Introduction}
At the present time, there is significant research interest in developing real-time schemes for intelligent radio sensing \cite{Bilik2019,Mishra2019,Aydogdu2020,Hassanien2019}. In particular, so-called `cognitive' radar systems have been proposed for both military and commercial applications. These systems aim to meet increasing performance demands while making minimal assumptions about the physical environment. Conceptually, cognitive radar has been imagined and re-imagined by many authors \cite{Haykin2012,Bell2015,Greco2018,Martone2014,Martone2021,Gurbuz2019}. Herein, we do not attempt to provide a comprehensive definition, but simply recognize cognitive radar to be a radar system which incrementally learns transmission and reception strategies from repeated experience. On an exhaustive definition of learning from experience, we also will not dwell, but instead take the interpretation of statistical learning theory, in which the learner's goal is to infer something about a phenomenon of interest by observing random samples of events related to said phenomenon \cite[Section 1.1]{Hajek2019}. Here, we study the fundamental problem of adaptively selecting transmission waveforms from a finite library, while making few \emph{a priori} assumptions about the physical scene. Instead, characteristics of the scene must be gradually inferred based on information gathered by the radar. 

An important consideration for cognitive radar is the development of \emph{sample-efficient} learning algorithms, which drive the adaptive selection of transmit and receive parameters \cite{Thornton2021}. A well-known challenge associated with data-driven approaches is that machine learning models are often parameter-heavy and prone to ``overfit" to a particular problem setting \cite{bousquet2002stability,ying2019overview}. Additionally, many machine learning models must be extensively trained ``offline" and may not be suitable for real-time interaction. Tracking radars sequentially probe the surrounding environment and data arrives sequentially in real-time. Thus, online learning is particularly important for cognitive tracking radar. These limitations present a particular challenge for active tracking systems, which must select transmission parameters using a limited number of noisy observations. Moreover, these measurements arrive sequentially, and the radar must quickly adapt to changes in the scene on-the-fly. To address these problems, the present work proposes \emph{online meta-learning} \cite{Finn2019} as a means of real-time waveform selection. 

Meta-learning is a process by which a decision maker gradually extracts knowledge from a sequence of observed learning experiences, or `tasks' \cite{Baxter1998}. This high-level knowledge is used to facilitate efficient learning of new tasks. The set of possible learning tasks is assumed to have an underlying relationship, but the tasks themselves are not identical. The meta-learner's goal is to find an \emph{inductive bias} that is suitable for efficient learning over the entire set of tasks. The inductive bias corresponds to a set of assumptions that the learner uses to make predictions about new data, where both time and observations are limited. In the present work, we study an online, or sequential approach to meta-learning where the inductive bias is gradually inferred over time after each task is completed. Here, the \emph{base learning} process occurs within a track while the \emph{meta-learning} process occurs on a track-to-track basis. 

Deducing the value of unseen future data using the past is impossible without further assumptions \cite{vapnik2006estimation}. Thus, all data-driven approaches are inherently biased by said assumptions. Selecting an appropriate bias by hand is often cumbersome, as encoding domain knowledge into complex model classes is notoriously difficult. It is important to note that conventional learning approaches involve learning a \emph{new model} from scratch for each task. Such approaches do not exploit task similarity, and can be data inefficient. An approach alternative to meta-learning is learning a single model which performs well on all tasks within a given class. Unfortunately, such a model may not exist, or sufficient data and learning time may not be available. This is especially true for the current application of target tracking.  

Here, we consider a scenario in which a stationary monostatic radar engages in a sequence of target tracks. The radar may track multiple targets at a time, and wishes to select waveforms which allow for accurate estimates of the target's position in the presence of clutter, interference, and noise. This waveform-agile tracking scenario is posed as an online meta-learning problem. We cast the repeated waveform selection process as a linear contextual bandit problem, and use a posterior (Thompson) sampling algorithm \cite{Agrawal2013,Russo2018} to select waveforms. Posterior sampling approaches are known to provide both good theoretical and practical performance, but are sensitive to the choice of prior distribution, which is difficult to select by hand. Thus, we aim to gradually \emph{learn} a prior distribution which allows for efficient learning across a class of tracking instances. For a stationary radar platform, we expect the target and clutter types the radar experiences from track-to-track to exhibit learnable structure, due to similarity across target and clutter impulse responses. Thus, we expect experience to carry over from one track to another, allowing the radar to learn an effective bias for any tracking instance. 

In this formulation, the waveform selection process for each time-limited target track corresponds to a new learning instance or `task'. We consider a very general signal model which incorporates spatially extended target and clutter responses, meaning that returns may occupy multiple range cells. This \emph{matched-illumination} set-up is commonly studied in the waveform design literature \cite{Kay2007,Romero2011,Pillai2000} and is appropriate for modern radar systems, which are capable of transmitting increasingly wideband pulses. Each learning task is then specified by the target's physical trajectory, the target impulse response, clutter impulse response, and additive noise model. 


We argue that meta-learning has significant potential for performance improvements in tracking radar systems due to several aspects of the problem. Firstly, the radar is assumed to be embedded in a physical environment which remains relatively stationary from track-to-track. We expect there to be a high degree of similarity between tracking instances, inducing some learnable structure. However, each track is not identical, due to the new targets' trajectory and changes in the physical scene. Further, the number of observations during each target track is limited, which requires the radar to learn quickly and sequentially. Finally, the first few measurements of a track are of particular importance so that the track is not lost. Thus, learning a reasonable waveform selection strategy quickly will result in significantly fewer lost tracks. Here, we propose a practical online meta-learning scheme that drastically reduces the probability of losing a track across a variety of physical scenes.


\subsection{Related Work}
This work addresses the problem of learning optimal adaptive waveform selection strategies for a broad class of radar environments. Therefore, some notes on the various notions of optimality for radar waveforms are required. There is a rich history of literature on radar waveform design, and providing a comprehensive survey is beyond the scope of the present work. The interested reader can consult several excellent surveys on waveform diversity and design \cite{Calderbank2009,Blunt2016,Roberts2010,Benedetto2009,gini2012waveform} as well as the canonical book on radar signals by Levanon \cite{Levanon2004}. Quoting Levanon, ``the work (or art) of designing radar waveforms is based mostly on experience and expertise obtained through successive designs" \cite[Section 1.5]{Levanon2004}. Here, we seek to automate the process of gaining experience by way of meta-learning.

The general principle of waveform design is to utilize a transmit signal that maximizes information gained about the target while accounting for channel effects due to propagation loss, noise, and interference. The following papers have been influential, but the list is by no means exhaustive.

Bell \cite{Bell1993} proposed mutual information between the target impulse response and received signal as a waveform design criterion for estimating the stochastic parameters of extended targets. Kay \cite{Kay2007} addressed the problem of optimal signal design and detection for point targets in signal-dependent Gaussian distributed clutter. Pillai et al. \cite{Pillai2000} consider $\sinr$-based design of pulsed waveforms for the detection of extended targets in signal-dependent noise, importantly showing that chirp-like signals are often insufficient in the presence of low-pass dominant noise. However, an issue with both $\sinr$ and mutual information-based waveform design procedures is that closed-form solutions are often not possible and on-the-fly waveform design can be computationally burdensome. These and other issues are addressed by Romero et al. \cite{Romero2011}, which extend the mutual-information based design to the case of signal-dependent clutter. Some of the information-theoretic trade-offs in waveform design for extended target detection have been examined by Zhu et al. \cite{Zhu2017,Zhu2018}. Other works have focused on optimizing the transmitted code sequence and receive filter simultaneously using sophisticated optimization techniques \cite{naghsh2013doppler,karbasi2015robust,wu2017transmit}.

In a real-time target tracking system, it is generally preferable to forgo waveform design and instead select from a catalog of known waveforms. This is due to various resource constraints that make on-the-fly waveform design difficult \cite{Cochran2009}. The problem of waveform-agile tracking is surveyed in \cite{Sira2009}. Numerous approaches have followed the Fisher information perspective, introduced by Kershaw and Evans \cite{Kershaw1994,Kershaw1997,Sira2007,Savage2007}. Other approaches have focused on waveform optimization by estimating clutter statistics on-the-fly \cite{Zhang2013,Aubry2013}. However, the aforementioned waveform selection techniques become computationally burdensome when longer-term scheduling is considered.  

It has been known for some time that non-myopic scheduling techniques are effective for a diverse range of sensor management problems, which often require planning \cite{LaScala2005,Charlish2015}. However, computational concerns historically limited the practical applications of stochastic control processes in large-scale waveform selection problems. In recent years, computational and algorithmic advances have allowed for real-time adaptive transmission schemes based on dynamic programming \cite{Selvi2020}, deep reinforcement learning \cite{Thornton2020}, universal source coding \cite{thornton2022universal}, and more computationally feasible schemes based on multi-armed bandit and contextual bandit learning \cite{Thornton2021,Howard2021}. While these learning schemes have proven to be effective in environments with stationary dynamics, the major issues of generalization and sample-efficiency remained understudied. In Table \ref{table:ml}, we outline the relative advantages and drawbacks of the learning methods described in such works briefly.

\begin{table*}[]
	\centering
	\caption{Machine learning approaches for adaptive waveform selection}
	\scriptsize
	\begin{tabular}{|l|l|l|l|l|l|}
		\hline
		& \cite{Howard2021,hero2007foundations,krishnamurthy2001hidden}       & \cite{Thornton2021} &\cite{Selvi2020,Thornton2020,LaScala2005,Charlish2015}                                                                                                                                                                      &       \cite{thornton2022universal}                                                                                                             & Present Work                                                                                                  \\ \hline
		Formulation & Multi-Armed Bandit                                                                                          & Contextual Bandit                                                                                                                              & \begin{tabular}[c]{@{}l@{}}MDP/\\ POMDP\end{tabular}                                                                                                                                  & Context-Tree Weighting                                                                                                      & \begin{tabular}[c]{@{}l@{}}Meta-Learning a \\ Contextual Bandit\end{tabular}                            \\ \hline
		Advantages  & \begin{tabular}[c]{@{}l@{}}Low-Complexity/\\ Online/\\ Bounded regret\end{tabular}  & \begin{tabular}[c]{@{}l@{}}Low-Complexity/\\ Context aware/\\ Online/\\ Bounded Regret\end{tabular}                             & \begin{tabular}[c]{@{}l@{}}Accounts for\\ impact of radar's actions/\\ Bellman Optimality/\\ Scales to large action \\ spaces via deep learning\end{tabular} & \begin{tabular}[c]{@{}l@{}}Bellman Optimality/\\ Few a priori \\ assumptions/\\ Online\end{tabular}  & \begin{tabular}[c]{@{}l@{}}Low-Complexity/\\ Bounded Regret/\\ Generalization/\\ Online\end{tabular} \\ \hline
		Drawbacks   & \begin{tabular}[c]{@{}l@{}}No context awareness/\\ Converges to \\ single action\end{tabular} & \begin{tabular}[c]{@{}l@{}}Does not account for impact \\ of radar's actions/\\ Unknown\\ generalization performance\end{tabular} & \begin{tabular}[c]{@{}l@{}}May require \\ offline training/\\ Sample inefficient/\\ Generalization unknown\end{tabular}                                                             & \begin{tabular}[c]{@{}l@{}}Limited to low \\ dimension problems/\\ Poor generalization \end{tabular} & Assumes task similarity                                                                                   \\ \hline
	\end{tabular}
\label{table:ml}
\end{table*}

This paper proposes online meta-learning as a method for real-time waveform selection. Meta-learning is broadly defined as a method of learning from examples which leverages experience gained from previous interactions. This principle is based on the idea that learning multiple related tasks should improve performance \cite{Thrun1996,Baxter1997,Baxter1998,Amit2018,Finn2019}. The meta-learner's goal is to generalize from a finite set of observed tasks to arbitrary new tasks. In this present context of radar tracking, we consider each track to be a new task. Our assumption at the outset is that learning a waveform selection strategy for a given target track should give the radar prior knowledge about which waveforms may be appropriate to use for future target tracks. This is because of similarities in the target impulse response and clutter impulse response from track-to-track. 


Recently, meta-learning has been proposed for several wireless communication and networking applications \cite{Simeone2020,Park2021,Kalor2021,Hu2021,Jiang2021}, with the goal of reducing training overhead and complexity. This application is reasonable, as domain knowledge is available in many wireless applications, but is often difficult to encode this knowledge by hand. Meta-learning circumvents this challenge by learning an inductive bias automatically. Several contributions have focused on using meta-learning to train detectors with a limited number of training examples \cite{Park2021,Jiang2021}. Here, we focus exclusively on the problem of transmit waveform selection. The number of observations available for learning is inherently limited by the application of target tracking. Additionally, we assume that the radar is physically stationary, from which we may expect a great deal of similarity across tracking instances. Thus, we argue that the waveform-agile tracking problem is well-posed to benefit from meta-learning techniques.

\subsection{Our Contributions}
This work is the first known investigation of online meta-learning to a waveform selection problem, other than our preliminary research \cite{Thornton2022}. In the current paper, we extend \cite{Thornton2022} by providing additional details regarding the learning framework and present general signal and channel models, incorporating both extended targets and signal-dependent clutter. We also present an extended simulation study. 

More specifically, we make the following contributions to the known literature: \
\begin{itemize}
	\item[C1.] We develop a general framework for the meta-learning problem by defining class of finite-state representations for the waveform-agile target tracking problem. To develop a model for this class, we extend the notion of a finite-state target channel, which describes the radar propagation environment for a specific tracking instance, \cite[Ch. 3.2.3]{Bell1988} to an ensemble, or \emph{class} of such channels that can be used to specify a meta-learning problem. 
	
	\item[C2.] We identify the benefits and limitations of Bayesian learning for the application of radar tracking in more detail than previous work, and discuss the significant impact of prior misspecification on tracking performance. In particular, we note that prior misspecification can result in a significant probability of low SNR measurements early in a track, when a target is most likely to be lost.

	\item[C3.] We demonstrate that a meta-Thompson sampling algorithm can be applied to effectively learn a prior for a class of finite-state target channels (FSTCs) satisfying sufficient structure, namely that the class of channels can be parameterized by a vector, and the FSTC for each tracking instance can be thought of as independent samples coming from a fixed distribution. We argue that this is due to inherent similarities between radar tracking instances, such as common target material and size. 
	
	\item[C4.] We discuss some limitations of the proposed meta-learning approach and discuss practical considerations. We identify how a meta-learning algorithm may interact with other processes in a multi-function radar system.
\end{itemize}

\subsection{Organization}
The remaining discussion is structured as follows. Section \ref{se:system} describes the radar signal model and propagation characteristics and proceeds to a discussion of the finite-state target channel considered. Section \ref{se:learning} presents the Bayesian learning problem of interest and describes the process of meta-learning by which the radar learns an inductive bias. Section \ref{se:algo} presents the meta-learning algorithm used and describes associated consideration. Section \ref{se:sim} examines the performance of the proposed approach in simulation. Section \ref{se:concl} provides concluding remarks.

\section{System Model}
\label{se:system}
We consider a physically stationary and monostatic radar system. The radar engages in an ongoing sequence of target tracks, in which the number of targets tracked and object trajectories may vary from one track to another. We assume the radar tracks one target at a time, and the target is physically extended in range against an extended clutter background. Each target track lasts for a fixed number $1 < n < \infty$ of Coherent Processing Intervals (CPIs), which we refer to as a tracking interval. During each tracking interval, the radar monitors a single target that is assumed to have been detected during an earlier scan period. The radar wishes to maintain adequate detection performance while accurately estimating target parameters of interest.

Let the area covered by the radar's beam be represented by a discrete sampling grid in delay-Doppler space, with delay cells indexed by $\tau = 1,...,\tau_{\max}$ and Doppler cells indexed by $\nu = 1,...,\nu_{\max}$. During each CPI, the radar transmits a finite-duration bandpass signal expressed by the model
\begin{equation}
	s_{T}(t) = \sqrt{2} \mathrm{Re}[\sqrt{E_{T}}\tilde{s}(t)\exp(j 2 \pi f_{c}t)],
\end{equation}
where $f_{c}$ is the carrier frequency, $E_{T}$ is the energy of the transmitted pulse, and $\tilde{s}(t)$ is the \emph{complex envelope} of the signal \cite[Appendix]{VanTrees2001}, which is assumed to be normalized, ie. $\int |\tilde{s}(t)|^{2} dt = 1$. We note that the delay-Doppler ambiguity properties of the transmitted waveform are completely determined by the complex envelope of the signal \cite[Chapter 3]{Levanon2004}. We assume that $E_{T}$ and $f_{c}$ are fixed, and wish only to vary the complex envelope on a CPI-to-CPI basis. This is a standard problem considered in the waveform design literature \cite{Kay2007,Aubry2013,Romero2011}. 

Here we assume the transmitted signal during each CPI is a coherent train of $N_{p}$ identical pulses. Thus, the complex envelope is expressed by
\begin{equation}
	\tilde{s}(t) = \frac{1}{\sqrt{N_{P}}} \sum_{i=1}^{N_{p}} p_{i}(t-(i-1)T_{r}),
\end{equation}
where $p_{i}$ is the complex envelope of the $i^{\text{th}}$ pulse in the CPI and $T_{r}$ is the pulse repetition period. 

In this formulation, we constrain the choice of envelope to a finite set of known waveforms. However, we wish to diversify the envelope enough such that a good waveform can be selected for a variety of tracking instances. Thus, we allow $p(t)$ to be either a generalized frequency modulated (GFM) chirp or a phase-coded pulse. The complex envelope of the GFM chirp is given by
\begin{equation}
	p(t) = a(t) \exp \left(j 2 \pi b \xi (t/t^{\prime}) \right),
\end{equation}
where $a(t)$ is an envelope function, $b$ is a scalar FM rate parameter, $\xi(\cdot)$ is the \emph{chirp phase function} \cite{Sira2007}, and $t^{\prime} > 0$ is a reference time point. The complex envelope of the phase-coded pulse is given by
\begin{equation}
	p(t) = \frac{1}{\sqrt{T}} \sum_{m=1}^{M} u_{m} \operatorname{rect} \left[ \frac{t-(m-1) t_{b}}{t_{b}} \right],
\end{equation}
where $M$ is the number of sub-pulses, $T$ is the pulse duration, $u_{m} = \exp(j \phi_{m})$, and the set of $M$ phases $\{\phi_{i}\}_{i=1}^{M}$ is the specific \emph{phase code} associated with $\tilde{s}(t)$. A detailed description of commonly used phase codes and their desirable properties can be found in \cite[Chapter 6]{Levanon2004}.

Each CPI, we assume that the transmitted pulse train illuminates both moving targets and surrounding clutter. The received signal is then given by 
\begin{equation}
	r(t) = \chi(t) * [s_{R}(t) + n(t)],
\end{equation}
where $\chi(t)$ is the complex-valued receive filter impulse response, which we take to be a matched filter (i.e. $\chi(t) = \tilde{s}^{*}(-t)$) here, $n(t)$ is a complex-valued \emph{signal-independent} additive noise term, due to receiver noise, interference or jamming. Finally, $s_{R}(t)$ is the reflected signal. Assuming $L$ targets and $H$ clutter components, the received signal during the CPI can be expressed as 
\begin{multline}
	s_{R}(t) = \sum_{g = 1}^{G} e^{j 2 \pi f^{g}_{d} t} \int_{0}^{\tau_{\max}^{g}} \frac{1}{R_{g}^{2}} \kappa_{g}(\tau) s \left(t- \left(\frac{R_{g}}{c}+\tau \right) \right) d \tau \\  + 	
	\sum_{h=1}^{H} e^{j 2 \pi f^{h}_{d} t} \int_{0}^{\tau_{\max}^{h}} \frac{1}{R_{h}^{2}} \lambda_{h}(\tau) s \left(t- \left(\frac{R_{h}}{c}+\tau \right) \right) d \tau
	\label{eq:reflected}
\end{multline}
where $\kappa_{g}(t)$ is the complex impulse response of target $g$, which may be modeled as a finite-energy stochastic process that varies on a CPI-to-CPI basis, $R_{g}$ and $f_{d,g}$ are the range Doppler frequency of target $g$ respectively, and $\lambda(t)$ is the clutter's complex impulse response, which also may be modeled as a finite-energy random process in general. We assume that the number of targets $L$ is known \emph{a priori} from a detection process.

Analogously, it is often more convenient to specify the target and clutter models by their frequency responses, given by
\begin{align}
	h_{g}(f) &= \int_{0}^{\tau_{\max}} \kappa_{g}(\tau) \exp(-j 2 \pi f \tau) d \tau \\
	c_{h}(f) &= \int_{0}^{\tau_{\max}} \lambda_{h}(\tau) \exp(-j 2 \pi f \tau) d \tau.
\end{align}

This system model assumes that both the target and clutter are \emph{extended} in range and thus exhibit frequency selectivity \cite[Chapter 12]{VanTrees2001}. This is a reasonable assumption when the target and clutter occupy multiple range cells, which is relevant for many high-resolution applications. For example, extended target modeling for missile and airplane shaped targets has been conducted in \cite{Kashyap1995,Cuomo1999,Jiang2001}. 

From (\ref{eq:reflected}), we observe that the complex envelope should in some sense be ``matched" to a composite channel response that incorporates both the target and clutter impulse responses. This is the problem investigated in \cite{Romero2011,deng2012waveform,leshem2007information}. However, in many cases it is difficult or impossible to solve for the optimal waveform in closed-form. Further, the approach of \cite{Romero2011,deng2012waveform} assumes knowledge of the target and clutter impulse response a priori, which is not practical for real-time tracking systems.

In our model (\ref{eq:reflected}), target illumination, clutter illumination, and various signal independent noise components all affect the received signal, and may change significantly over the course of a track. Thus, solving for the optimal waveform in closed-form will be intractable in general. A helpful model for such time-varying channels is the finite-state representation \cite[Section 4.6]{Gallager1968}. This model has been widely applied in communication theory to model channels with intersymbol interference and fading, and has also been suggested for radar \cite{Bell1988}. 

To motivate the meta-learning problem, we consider a scenario in which each tracking instance is described by a unique finite-state target channel (FSTC) with memory. We assume that the radar engages in an ongoing sequence of target tracks, and that there is an underlying \emph{class} of FSTCs that compose the physical scene, accounting for variability due to changes in the target and clutter impulse responses. In the next section, we introduce the FSTC model and relevant assumptions.

\subsection{Finite State Target Channel Model}
\label{se:fstc}
Each discrete time index\footnote{For convenience, we consider discrete time with finite observation and waveform sets. The extension to continuous domains comes with a loss of generality, and requires more sophisticated assumptions.} denoted by $k=1,2,...,n$ corresponds to the present radar CPI within a target track. During CPI $k$, the radar scene is said to be in a \emph{state} denoted by $s_{k} $, which\footnote{We use the notation $s_{1}^{k}$ to denote the sequence of states from time $1$ to $k$.} takes values in finite set $\mathcal{S}$. The state describes relative losses due to the scattering effects of the propagation environment, as well as from the target, representing a general notion of the channel's effect. Thus, the state captures the relevant effects present in (\ref{eq:reflected}), namely due to the targets' positions, Doppler shifts, and impulse responses, as well as the analogous scatterers modeled as clutter.

The state varies over time according to a discrete-time stochastic process $\{s_{k}\}_{k \in \mathbb{N}_{+}}$, with a fixed memory length $L < \infty$. More specifically, this means that knowledge of the entire state sequence does not provide more information about the current state than knowledge of the past $L$ states. In mathematical terms,
\begin{equation}
	P(s_{k}|s_{1}^{k-1}) = P(s_{k}|s_{k-L+1}^{k-1}),
	\label{eq:tProbs}
\end{equation} 
for every possible state sequence. The memory length is unknown to the radar. Thus, the state generating process is said to have transition probabilities $P(s_{k}|s_{k-L+1}^{k-1})$, which are unknown to the radar \emph{a priori}. In this formulation, we assume the scene's state transitions occur independent of the radar's transmissions. This is true of most radar tracking scenarios, but notably excludes the cases of an adversarial target response or adversarial interference.

Since the radar is a measurement system, it does not observe the underlying state directly. The radar instead has access to an observation $o_{k} \in \mathcal{O}$ at each time step. This observation is a hypothesis regarding the target's position and the scattering effects of the channel. The observation process is governed by the probability kernel $P(o_{k}|s_{k})$, defined for each $o_{k} \in \ncalO$ and $s_{k} \in \ncalS$. 

Based on the sequence of observations $o_{1}^{k}$, the radar must select a waveform $w_{k}$ from a finite alphabet of waveforms $\mathcal{W}$. The radar wishes to measure a random vector of target parameters $z_{k} \in \mathcal{Z}$, generated by a fixed source distribution $P(z_{k})$, which is assumed to be independent of the radar's waveform choice and the state of the scene. We have now established the tools necessary to define our finite state target channel (FSTC), which the radar will repeatedly interact with during a target track.

\begin{ndef}[FSTC]
	A FSTC is specified by the state transition probabilities defined in (\ref{eq:tProbs}), the source distribution $P(z_{k})$, and the tuple 
	\begin{equation*}
		\left( \underbrace{\ncalW \times \ncalS \times \ncalZ}_{\text{Channel Inputs}}, \; \underbrace{P(y_{k}|w_{k},s_{k},z_{k})}_{\text{Channel Law}}, \underbrace{\ncalY}_{\text{Channel Output}} \right), 
	\end{equation*}
	where $\ncalS$ is a finite alphabet of states, $\ncalW$ is a finite waveform alphabet, $\ncalZ$ is a finite set of possible target parameter vectors, $\ncalY$ is a not necessarily finite set of received signals, and $P: \ncalW \times \ncalS \times \ncalZ \mapsto \ncalY$ matrix of probabilities which expresses the probability of receiving a particular signal given a waveform, state sequence, and target parameter.
	\label{def:fstc}
\end{ndef}

\begin{remark}
	Our channel model is a generalization of the FSTC proposed by Bell \cite{Bell1988}, as the state sequence of our FSTC is not assumed to be a $1^{\text{st}}$ order Markov process, and may depend on the entire past state sequence. Our FSTC also differs from that of Bell in that we do not allow the target parameter vector to depend on the radar's choice of waveform. 
\end{remark}

\begin{remark}
	A conventional learning approach would seek to optimize the choice of waveforms for a fixed, but a priori unknown, FSTC. We will refer to the problem of learning a waveform selection strategy for a particular FSTC as the \emph{task}, to remain consistent with the meta-learning literature. Meta-learning aims to find an inductive bias which allows for efficient learning across a \textbf{class} of FSTCs with parameters drawn from a common distribution.
\end{remark}

\begin{ndef}[FSTC Class]
	Given a common state, waveform, and target parameter alphabet $(\ncalW, \ncalS, \ncalZ)$, we can consider a class $\Theta$ of FSTCs, where each channel is specified by the parameter vector $\theta \in \Theta$ and is characterized by the conditional probability assignment
	\begin{equation}
		P(y|w,s,z,\theta),
	\end{equation}
	for every $w \in \ncalW$, $s \in \ncalS$, and $z \in \ncalZ$.
\end{ndef}

The above definition is analogous to a class of finite-state fading channels \cite{Lapidoth1998}. The parameter vector $\theta$ specifies the unique propagation characteristics of each channel not captured by the state. These effects are due to any variations in the target and clutter impulse responses in (\ref{eq:reflected}) which are not explicitly captured by the state. It is this parameter vector which we seek to learn in Section \ref{se:learning}. 

\begin{assum}[Task Structure]
	Each target track corresponds to a new FSTC, drawn independently from a fixed class $\Theta$. For a sequence of tracks, we assume the existence of a distribution $P(\theta)$ on $\Theta$. This assumption is reasonable since a new target track corresponds to a new trajectory, target scattering behavior, and possible changes in clutter and noise distribution. However, since the physical scene is expected to remain relatively stationary over time, the scene should have enough structure to define $\Theta$ and $P(\theta)$.
\end{assum}

Since capacity-achieving schemes are impractical without explicit assumptions and \emph{a priori} knowledge of the target channel's behavior, we frame the waveform selection process for a fixed FSTC as a statistical learning problem. During each track instance, the radar wishes to select the waveform which minimizes an unknown loss function during each CPI, which is a sequential decision problem over a finite time horizon. The loss represents a performance characteristic which is calculable without knowledge of the underlying channel state. Instead of learning a fixed waveform selection strategy that performs well across an entire class of FSTCs, we seek to learn an inductive bias that allows for efficient learning of any new channel within the class. 

\section{Learning Framework}
\label{se:learning}
The radar engages in a sequence of $m$ target tracking periods. During each tracking instance, $s = \{1,2,...,m\}$, the radar maintains an estimate of the target parameter vector $\{z_{k} \}_{k=1}^{n}$ over a fixed time horizon of $n$ CPIs. Based on the selected waveform at CPI $k$, namely $w_{k}$ and underlying state\footnote{As mentioned in Section \ref{se:fstc}, the true value of the state is unobserved by the radar} $s_{k}$ the radar receives a loss\footnote{We describe our specific choice of loss function in Section \ref{se:sim}. An overview of possible choices can be found in \cite[Section III.]{charlish2020implementing}.}, defined by the sequence of mappings $\ell_{k}: \ncalW \times \ncalS \mapsto \mathbb{R}$. The loss must be a quantity which is be directly observable by the radar. For example, features from the range-Doppler response, such as estimated SINR, or outputs from the radar's tracking filter, such as the normalized innovation, are natural choices for the loss function \cite[Section III.]{charlish2020implementing}. Since the state is not fully observable by the radar, the loss appears stochastic. The goal of the learning problem at each time step is to select the waveform
\begin{equation}
	 w_{k}^{*} = \underset{w_{k} \in \ncalW}{\min} \E[\ell_{k}(w_{k},s_{k})], 
\end{equation} 
where the expectation is taken over the variability in the loss mapping. Unfortunately, this expectation is not computable as both the sequences of loss mappings $\{\ell_{k}\}$ and state values $\{s_{k}\}$ are unknown to the radar. Thus, the radar must select waveforms using the only the history of observations, transmitted waveforms, and received losses up to step $k-1$, given by the set
\begin{equation}
	\ncalF_{k-1} = \{(o_{i},w_{i},\ell_{i})\}_{i=1}^{k-1}.
\end{equation} 

In general, this problem is practically challenging for two reasons. 
\begin{enumerate}
	\item Each state value $s_{k}$ is only partially observed through $o_{k}$. Thus, the radar must select waveforms with only partial knowledge of the underlying FSTC.
	\item The structure of $\ell_{k}$ is unknown to the radar \emph{a priori} and must be inferred in the small-sample regime. Thus, the radar must gather information about the losses associated with each waveform, while trying to minimize the total number of sub-optimal waveforms transmitted.
\end{enumerate}\vspace{.2cm}

Thus, the radar must gather information about the current FSTC instantiation by gradually building $\ncalF_{k-1}$ through repeated experience. The information gathered in the history must be sufficiently rich, such that the radar has enough knowledge to select near-optimal waveforms. However, this information must be gathered quickly, such that the total number of sub-optimal waveforms transmitted during the tracking interval is kept low. Thus, the canonical dilemma of \emph{exploration} and \emph{exploitation} is present. This challenge is particularly relevant for the inherently time-sensitive tracking problem, as the number of sub-optimal waveforms transmitted must be kept low throughout the entirety of the tracking instance to avoid losing the target. 



We now introduce a pseudo-Bayesian learning formulation, in which waveforms are selected based on the estimated posterior probability that they are optimal. We develop a scheme to estimate the FSTC parameter $\theta$, which can be used to map radar observations to expected losses. We note that in this scheme, the estimated posterior distribution does not model any inherent randomness in the value of $\theta$ itself, but rather reflects the radar's uncertainty about the true value of $\theta$.

In essence, the we parameterize the present target track by a vector $\theta \in \nbbR^{d}$, which can be used to map the radar's observations to expected losses. We express uncertainty about the value of $\theta$ by introducing a prior distribution $P(\theta)$. As the radar gains experience, the posterior distribution $P(\theta|\ncalF)$ begins to concentrate around a certain value. We have thus reduced the sequential decision problem to a density estimation problem. This approach, known as posterior sampling, or \emph{Thompson sampling}\footnote{Due to the foundational work of Thompson \cite{Thompson1933}.}, is widely used for sequential decision making due to a unique combination of theoretical performance and practical efficiency \cite{Lattimore2020,Russo2018}. 


\subsection{Thompson Sampling Approach}
A common framework used for sequential decision making is the Bayesian approach. Under this framework, decisions are made with respect to the posterior probability that they are optimal \cite[Section 2.2]{Baxter1998}. This approach assumes the learning task can be described by a set of probability distributions $\{P_{\theta}\}$ parameterized by $\theta \in \Theta \subset \nbbR^{d}$.

An computationally feasible example of Bayesian learning is posterior sampling in the well-known \emph{stochastic linear bandit} problem \cite[Ch. 19]{Lattimore2020}, which is adopted here. In this framework, we define $\varphi: \ncalO \times \ncalW \mapsto \nbbR^{d}$ to be a feature mapping\footnote{We discuss considerations related to our specific choice of feature mappings in Section \ref{se:sim}.}, which describes aspects of the physical scene. The stochastic linear bandit model is then specified by assuming the relation
\begin{equation}
	\label{eq:innerProd}
	\ell_{k} = \langle \theta^{*}, \varphi(o_{i},w_{i}) \rangle + \eta_{k},
\end{equation}
holds for each pair $(o_{i},w_{i}) \in \ncalO \times \ncalW$, where $\eta_{k}$ is a sub-Gaussian random variable\footnote{A random variable $\xi$ is sub-Gaussian if, for all $\lambda \in \nbbR$, it holds that $\E[\exp(\lambda \xi)] \leq \exp(\lambda^{2}/2)$.}. Under this assumption, knowledge of the vector $\theta$ is sufficient to predict $\E[\ell_{k}]$ for each pair of waveform and observation $(o_{k},w_{k})$. Taking a Bayeisan perspective, the radar must use observed data $\ncalF$ to learn a posterior distribution over the parameter $P(\theta|\ncalF)$. This is the base learning problem we will study in the remainder of this paper.

Given the stochastic linear bandit assumption (\ref{eq:innerProd}), the ability to parameterize the FSTC by $\theta$ as discussed in Definition \ref{def:fstc} is of great importance, and we now discuss it briefly. The inner product relation (\ref{eq:innerProd}) dictates the mapping of context features onto expected losses with a reasonable degree of certainty. In order for such a relation to be satisfied for a realistic tracking scenario, the FSTC must be adequately described by the set of features $\{\varphi(o_{i},w_{j})\} \quad \forall i \in \ncalO, j \in \ncalW$. The function $\varphi$ maps observable quantities, such as estimated target position and $\sinr$, to a vector describing the utility of waveform $w_{i}$. Additionally, a suitable weighting over these features must reliably predict the associated loss for each pair $(o,w)$.  

We now describe the process for estimating a posterior distribution describing the radar's uncertainty about the value of $\theta$ using the radar's observations directly. The radar starts with a prior over possible FSTC parameter values $P(\theta)$. Using the observed data $\ncalF_{k-1}$, the radar then updates its prior distribution to a posterior distribution by applying Bayes rule
\begin{equation}
	P(\theta|\mathcal{F}_{k-1}) = \frac{P(\mathcal{F}_{k-1}|\theta)P(\theta)}{P(\mathcal{F}_{k-1})} \propto P(\mathcal{F}_{k-1}|\theta)P(\theta).
	\label{eq:posterior}
\end{equation}
Since $\theta$ is a fixed, continuous-valued parameter vector, we may reflect the radar's uncertainty about its value by choosing a Gaussian prior distribution. Assuming a Gaussian likelihood function, computation of the posterior in (\ref{eq:posterior}) is straightforward and computationally efficient. A detailed computation of (\ref{eq:posterior}) may be found in Appendix \ref{se:post}.

\begin{ndef}[Thompson Sampling]
	Thompson sampling proceeds by selecting the waveform $w_{\text{TS}} \in \ncalW$ such that 
	\begin{equation}
		\label{eq:tspolicy}
		\int_{\Theta} \mathbbm{1} \left[ \E[\ell|w_{\text{TS}},o,\theta] = \min_{w'} \E[\ell|w',o,\theta] \right] P(\theta|\ncalF_{k-1}) d\theta,
	\end{equation}
	where $\mathbbm{1}[\cdot]$ is the indicator function. Details of a practical algorithm to achieve this policy in the linear contextual bandit setting can be found in \cite{Agrawal2013}. Details of a context-aware waveform selection strategy based on Thompson Sampling can be found in \cite{Thornton2021}.
\end{ndef}

\begin{ndef}[Bayesian Regret]
	The expected, or Bayesian, $n$-round regret of a decision strategy is given by
	\begin{equation}
		\operatorname{BR}^{*}_{n} \triangleq \E \left[\sum_{k=1}^{n} \ell_{k} - \sum_{k=1}^{n} \min_{w^{*} \in \mathcal{W}} \langle \theta, w^{*} \rangle \right],
	\end{equation}
	which is an indication of the average performance of a decision algorithm. However, an algorithm with low Bayesian regret may perform poorly in particular scenarios.
\end{ndef}

\begin{remark}
	While many frequentist bounds on the performance of TS exist \cite{Gopalan2014thompson,Agrawal2013}, they generally focus on the asymptotic case, where the effects of the prior distribution ``wash out". In practice, performance can be highly suboptimal if the prior is misspecified and the number of observations is limited, as in a target tracking problem.
\end{remark}

\begin{theorem}[Prior Dependent Upper/Lower Bound for Bayesian Bandit \cite{Lattimore2020}]
	For any prior distribution $P$ over the parameter space $\Theta$, the Bayesian regret of an $A$-armed bandit satisfies
	\begin{equation}
		\operatorname{BR}^{*}_{n}(P) \leq C \sqrt{A \times n},
	\end{equation}
	where $C > 0$ is a universal constant. Further, there exists a prior $P$ such that
	\begin{equation}
		\operatorname{BR}^{*}_{n}(P) \geq c \sqrt{A \times n},
	\end{equation}
	where $c > 0$ is a universal constant. Thus, we see that $\sup_{P} \operatorname{BR}^{*}_{n}(P) = \Theta(\sqrt{A \times n})$, where $f(x) = \Theta(g(x))$ denotes that $f(x)$ is bounded both above and below by $g(x)$ asymptotically.
\end{theorem}

Thus, we see that for the case of small time-horizon $n$, the choice of prior can make a relatively large difference. For example, in \cite{Bubeck2013}, a lower bound on the Bayesian regret of TS is established to be $\frac{1}{20}\sqrt{A \times n}$, along with a prior independent upper-bound of $14\sqrt{A \times n}$. 

Since the empirical performance of TS over a limited time horizon is highly dependent on the choice of prior distribution \cite{Bubeck2013,Dudik2021}, we discuss a meta-learning procedure for sequentially learning a prior over time, when the radar is embedded in an environment of tasks coming from a common \emph{task distribution}. We will see that in many scenarios, meta-learning an effective prior for Thompson Sampling can provide major benefits in tracking performance. We now describe a meta-learning formulation in which uncertainty about the prior distribution is captured by a modeling a distribution over instance priors. 

\begin{theorem}[Single-Task PAC-Bayes Bound \cite{mcallester1999pac}]
	Let $P \in \ncalM$ be a prior distribution over the hypothesis space $\Theta$. Then for any $\delta \in (0,1]$, the following inequality holds uniformly for all posterior distributions $Q \in \ncalM$ with probability at least $1-\delta$,
	\begin{equation}
		\operatorname{er}(Q, \mathcal{D}) \leq \underbrace{\widehat{e r}(Q, S)}_{\text{Empirical Error}}+ \underbrace{\sqrt{\frac{D_{\texttt{KL}}(Q \| P)+\log \frac{m_{s}}{\delta}}{2(m_{s}-1)}}}_{\text{Complexity term}},
	\end{equation}
	where $\operatorname{er}(Q, \mathcal{D})$ is the expected loss when acting according to posterior distribution $Q$ when the true underlying distribution is $\mathcal{D}$, $\widehat{e r}(Q, S)$ is the empirically observed loss when data $S \sim \mathcal{D}^{m_{s}}$ is observed, and $m_{s}$ is the sample size.
\end{theorem}

\begin{remark}
	While appearing abstract, the two terms in the PAC-Bayes bound illustrate a trade-off between fitting the data and complexity. The first term encourages picking a $Q$ that provides low empirical error, while the second term emphasizes $Q$ with a small distance from the prior $P$. The choice of prior affects the tightness of the bound. In general, we want to select a prior which is close to posteriors which give us low empirical error.
\end{remark}

\begin{algorithm}[t]
	\setlength{\textfloatsep}{0pt}
	\label{algo:ts}
	\caption{Adaptive Radar Waveform Selection via Thompson Sampling}
	\SetAlgoLined
	\textbf{Input} prior distribution $\ncalN(\hat{\theta},B^{-1})$, Loss function $\ell$\\
	Set $Q_{1} \leftarrow \ncalN(\hat{\theta},B^{-1})$ \\
	\For{\text{Each radar CPI $k= \; 1,...,n$}}{
		\vspace{0.07cm}
		(1) Sample weighting vector $\theta \sim Q_{k}(\theta|\ncalF_{k-1})$;\\ \vspace{.2cm}
		(2) Radar specifies observation of current target and channel state $o_{k}$;\\ \vspace{0.2cm}
		(3) Assemble context vectors $\{\varphi_{i}\}$ for each $w_{i} \in \ncalW$;\\ \vspace{0.2cm}
		(4) Take inner product $\lA \varphi_{i}, \theta \rA$ for each $w_{i} \in \ncalW$; \\ \vspace{0.2cm}
		(5) Select waveform $w_{k}^{*} = \underset{w_{i} \in \ncalW} \max \lA \varphi_{i}, \theta \rA$; \\ \vspace{.2cm}
		(6) Update history $\ncalF_{k}$ and posterior parameters $Q_{k+1}$;
	}
\end{algorithm}

\subsection{Meta-Learning Formulation}
In this section, we describe a meta-learning procedure which aims to improve radar tracking performance using information gained via repeated interaction over $m$ target tracking instances. Each individual tracking instance corresponds to a new instance of a $n$-step waveform-agile tracking problem on a FSTC described in Section \ref{se:fstc}, where the radar has access to a catalog of $|\ncalW| = K$ waveforms. 

The radar ultimately wishes to learn the true parameter vector for each tracking instance $s$, denoted by $\theta_{\star} \in \mathbb{R}^{d}$. Given knowledge of $\theta_{\star}$, contexts can be directly mapped to average losses using the inner product relationship $\ell_{k} = \langle \theta, \varphi(o_{k},w_{k}) \rangle + \eta_{k}$ for stochastic linear bandits. As candidates, we consider the class of distributions $\mathcal{P} = \{P_{\theta} : \theta \in \Theta \}$, where $\Theta$ is a compact subset of $\mathbb{R}^{d}$. Just as in the conventional learning setting, the radar is equipped with a prior distribution $P(\theta)$ which expresses a subjective belief about the value of $\theta$ based on the current sequence of observations $\ncalF$.

At the beginning of each tracking instance $v \in \{1,2,...N_{\text{track}}\}$, an instance of the learning problem, specified by $\theta_{v,\star}$ is sampled from a \emph{task distribution} $P_{\star}$, which is fixed but unknown to the radar. Since $P_{\star}$ corresponds to the true distribution of tasks, each task being parameterized by $\theta_{v,\star}$, it can also be interpreted as the best possible choice of prior for a Bayesian learning algorithm. Thus, the meta-learning process consists of estimating the true prior $P_{\star}$ by sequentially interacting with problem instances that are assumed to be sampled i.i.d from a common distribution $Q$. 

The radar's uncertainty about the true value of $P_{\star}$ is reflected by assuming it is sampled from a fixed distribution over instance priors $P_{\star} \sim Q$. We refer to $Q$ as a \emph{meta-prior}\footnote{This quantity is sometimes referred to as the \emph{hyper-prior} in the field of Bayesian statistics.}. The meta-learning environment is then characterized by the pair $(\ncalP,Q)$. We denote by $Q_{v}$ the \emph{meta-posterior}, which is the radar's current estimate of the instance prior $P_{\star}$ using information gained up to track $v$. Good performance will occur when $D_{\texttt{KL}}(Q_{v}||P_{\star})$ is small, which corresponds to a low degree of uncertainty regarding the true prior $P_{\star}$. The meta-posterior is estimated sequentially using the standard Bayesian update rule
\begin{align}
	&Q_{v+1}(\hat{P}) \propto P(H_{v}|P_{\star} = \hat{P}) Q_{v}(\hat{P}) \nonumber \\ &= Q_{v}(\hat{P}) \int_{\theta} P(H_{v}|\theta_{v,\star} = \theta) P(\theta_{v,\star} = \theta | P_{\star} = \hat{P}) d\theta.
	\label{eq:metaUpdate}
\end{align}

The meta-learning formulation is an example of a hierarchical Bayes model.

\section{Algorithm and Practice}
\label{se:algo}

\begin{algorithm}[t]
	\setlength{\textfloatsep}{0pt}
	\label{algo:mts}
	\caption{Contextual Meta-TS for Waveform Agile Tracking}
	\SetAlgoLined
	\textbf{Input} meta-prior distribution $Q$, Loss function $\ell$\\
	Set $Q_{1} \leftarrow Q$ \\
	\For{\text{Each target track $v = \; 1,...,N_{\text{track}}$}}{
		\vspace{0.07cm}
		(1) Sample $P_{v} \sim Q_{v}$;\\ \vspace{.2cm}
		(2) Apply Thompson Sampling waveform selection policy (\ref{eq:tspolicy}) with prior $P_{v}$ to problem parameterized by $\theta_{v,\star} \sim P_{\star}$ for $n$ CPIs;\\ \vspace{0.2cm}
		(3) Update Meta-Posterior $Q_{v+1}$ according to the update rules (\ref{eq:postupdateS}) and (\ref{eq:postupdate}); \\
	}
\end{algorithm}

At the heart of the waveform selection approach is posterior sampling \cite{Agrawal2013,Thornton2021}, which is also described for the waveform-agile target tracking problem in Algorithm \ref{algo:ts}. The meta-learning algorithm used in this paper is described in Algorithm \ref{algo:mts}, which seeks to learn an effective prior distribution for the linear contextual bandit described in Algorithm \ref{algo:ts}.  

To ensure that the meta-posterior update (\ref{eq:metaUpdate}) is computable in closed-form, and the existence of an efficient algorithm for online meta-learning, we consider a normal-normal conjugacy scenario, in which we study a Gaussian bandit with a Gaussian meta-prior. The Gaussian-Gaussian bandit model is commonly used in the literature on sequential decision processes over continuous parameter spaces. We note that the parameter $\theta$ is assumed to be a fixed value over each target track, and the Gaussian model is used here to reflect the radar's uncertainty about the true value of $\theta$. Thus, explicit assumptions about the environment are avoided. We assume a normal distribution over instance priors, expressed by $P(\theta) = \mathcal{N}(\mu,\sigma_{0}^{2}I_{d})$, where the noise level $\sigma_{0}^{2}$ is fixed. 

The meta-prior is then a distribution over instance prior means, given by $Q(\mu) = \mathcal{N}(\mathbf{0},\sigma_{q}^{2}I_{d})$. We assume the noise level $\sigma_{q}^{2}$ is fixed and known to the radar. The meta-learning process then maintains a meta-posterior $Q_{v}(\mu) = \ncalN(\hat{\mu}_{0,v},\hat{\Sigma}_{v})$. Due to the conjugacy properties of the normal distribution, the meta-posterior $Q_{v}$ can be simply updated in closed-form, even for a contextual bandit algorithm. Following the standard computations for Bayesian multi-task regression (reviewed in Appendix D of \cite{Kveton2021}), we see that the meta-posterior can be simply expressed by
\begin{equation}
	Q_{v} \sim \ncalN(\mu_{v},\Lambda_{v}^{-1}),
\end{equation}
where the parameters are updated each track $v$ by
\begin{align}
	\label{eq:postupdateS}
	\Lambda_{v} &= \Lambda_{v-1} + X_{v}^{T}(\sigma^{2}I+X_{v}\Sigma X_{v}^{T})^{-1} X_{v}, \\
	\mu_{v}    &= \Lambda_{v}^{-1}(\Lambda_{v-1}\mu_{v-1}+ X_{v}^{T}(\sigma I_{n}) + X_{v} \Sigma X_{v}^{T})^{-1} L_{v}),
	\label{eq:postupdate}
\end{align}
where $\Sigma = \sigma_{0}^{2}I_{d}$, $X_{v} = [\varphi_{k=1},\varphi_{2},...,\varphi_{n}]$ is the vector of observed contexts over each time index $k = 1,2,...,n$ of the present track $v$, and $L_{v} = [\ell_{k=1},\ell_{2},...,\ell_{n}]$ is a vector containing the observed losses during track $v$. The updates (\ref{eq:postupdateS}) and (\ref{eq:postupdate}) are easily calculated for low-dimension cases, such as the waveform selection problem examined here. For high-dimensional problems, the Woodbury matrix identity can be applied to speed up computations. A description of the meta-learning process can be seen in Algorithm \ref{algo:mts}.

We expect meta-learning to provide major benefits for cases where $\sigma_{q}^{2} \gg \sigma_{0}^{2}$, since uncertainty about the true prior $P_{\star}$ is significant. In cases where $\sigma_{q}^{2} \ll \sigma_{0}^{2}$, meta-learning is very close to the standard single-task learning problem, and we expect little benefit. Thus, the amount of relatedness between tasks is an important indicator of meta-learning performance. The issue of task similarity has been studied from an information-theoretic perspective in \cite{Jose2022}. Future work in this domain could focus on a rigorous justification of task similarity for radar tasks of practical interest, such as tracking under interference.

\begin{figure*}[t]
	\centering
	\subfloat[]{\includegraphics[scale=0.55]{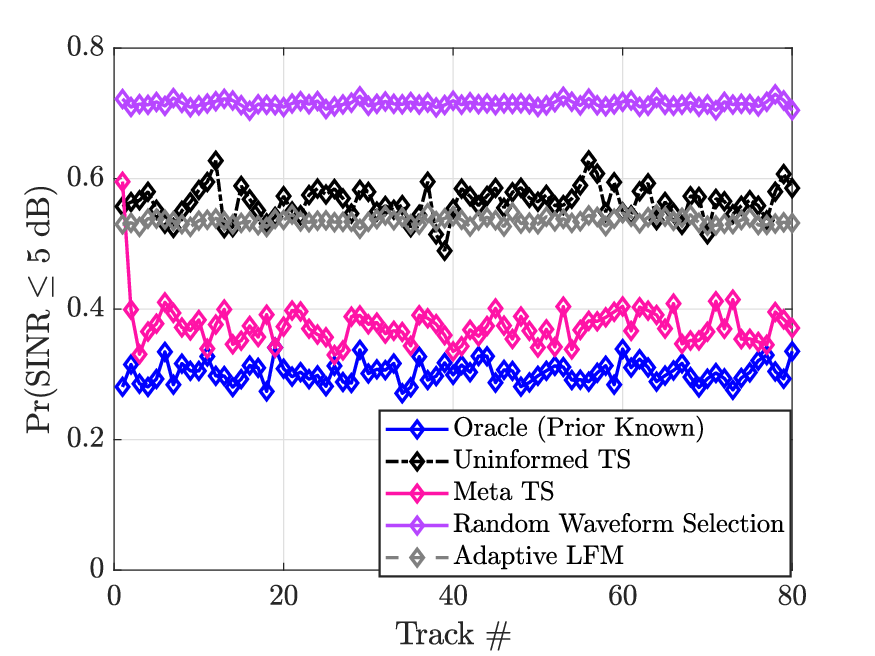}}
	\subfloat[]{\includegraphics[scale=0.55]{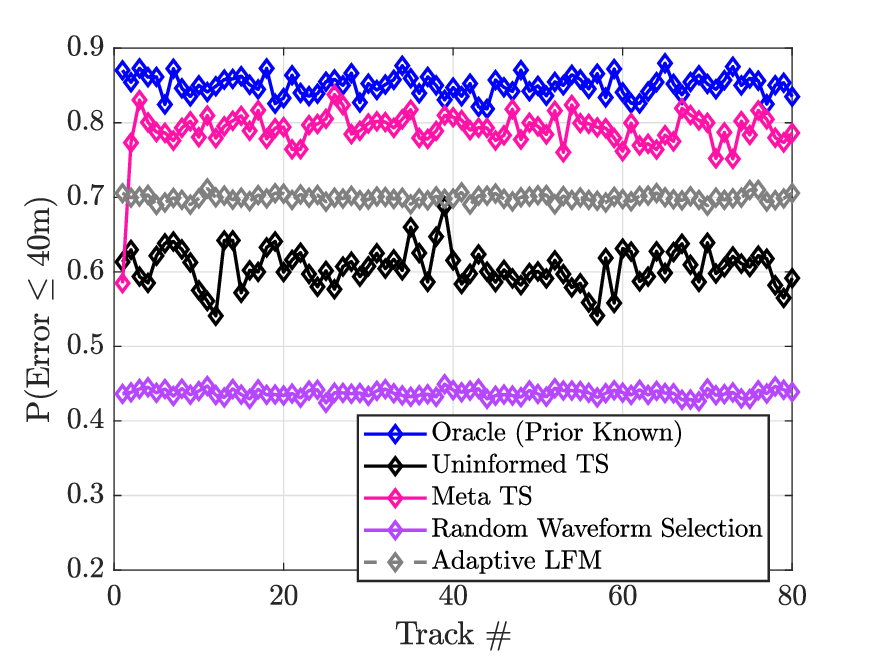}} \\
	\subfloat[]{\includegraphics[scale=0.55]{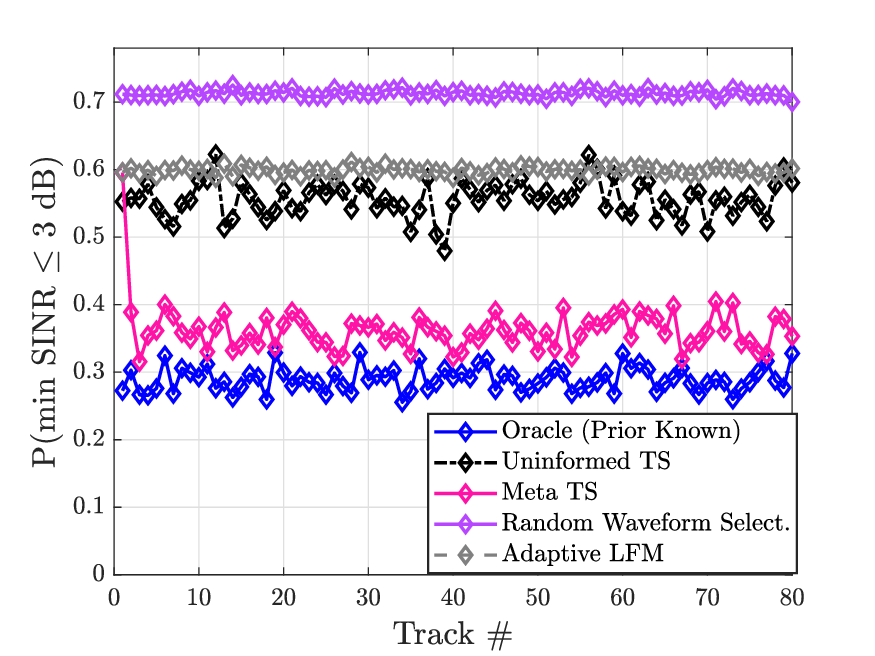}}
	\subfloat[]{\includegraphics[scale=0.55]{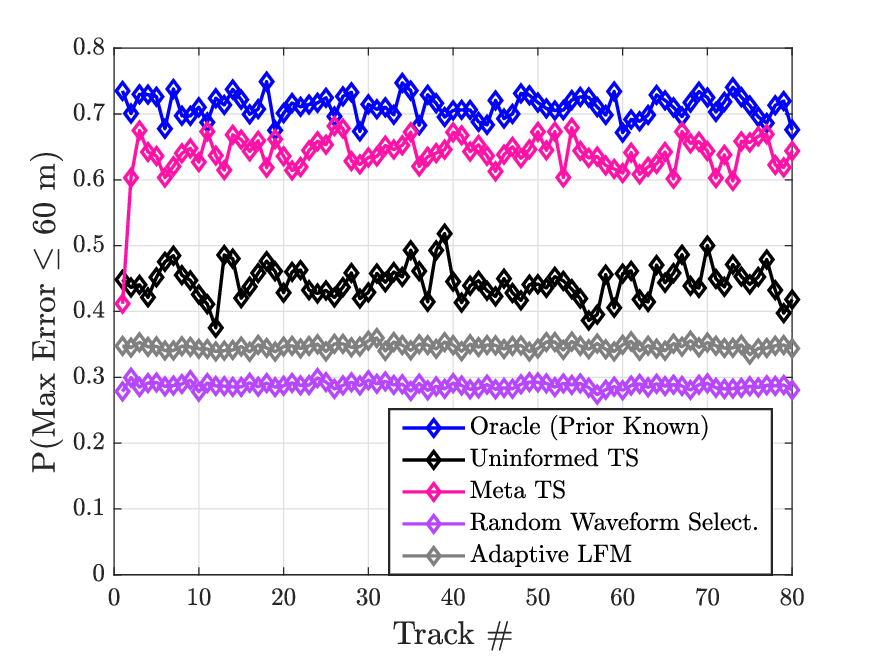}}
	\caption{\textsc{Outage analysis} in terms of (a) average target SINR, (b) average range tracking error, (c) minimum target SINR, and (d) maximum range tracking error.}
	\label{fig:outage_time}
\end{figure*}

\begin{theorem}[Extended PAC-Bayes Bound \cite{Amit2018}]
	Let $Q$ be a learning algorithm which maps observations and a prior to a posterior distribution on $\ncalM$, and let $\ncalP$ be some pre-defined hyper-prior distribution. Let $\tau_{v}$ be a new learning instance \cite[Sec. 3.1]{Amit2018}. Then for any $\delta \in (0,1]$, the following holds uniformly for the set of all hyper-posterior distributions $\ncalQ$ with probability at least $1-\delta$,
	\begin{multline}
		\operatorname{er}(\mathcal{Q}, \tau_{v}) \leq \underbrace{\frac{1}{n_{\text{obs}}} \sum_{i=1}^{n_{\text{obs}}} \underset{P \sim \mathcal{Q}}{\mathbb{E}} \widehat{e r}_{i}\left(Q_{i}, S_{i}\right)}_{\text{Empirical multi-task error}} \\
		+ \underbrace{\frac{1}{n_{\text{obs}}} \sum_{i=1}^{n_{\text{obs}}} \sqrt{\frac{D_{\texttt{KL}}(\mathcal{Q} \| \mathcal{P})+\underset{P \sim \mathcal{Q}}{\mathbb{E}} D_{\texttt{KL}}\left(Q_{i} \| P\right)+\log \frac{2 n_{\text{obs}} m_{i}}{\delta}}{2\left(m_{i}-1\right)}}}_{\text{Avg. Task Complexity}} \\
		+\underbrace{\sqrt{\frac{D_{\texttt{KL}}(\mathcal{Q} \| \mathcal{P})+\log \frac{2 n_{\text{obs}}}{\delta}}{2(n_{\text{obs}}-1)}}}_{\text{Environment Complexity}},
	\end{multline}
	where $Q_{i} \triangleq Q(S_{i},P)$.
\end{theorem}

\begin{remark}
	The extended PAC-Bayes bound consists of the empirical multi-task error plus two complexity terms. The first is the average of task complexity terms for the observed tasks. This term tends to zero as the number of samples in each task $m_{i} \rightarrow \infty$.
\end{remark}

\section{Simulation Study}
\label{se:sim}

\begin{figure*}
	\centering
    \subfloat[]{\includegraphics[scale=0.55]{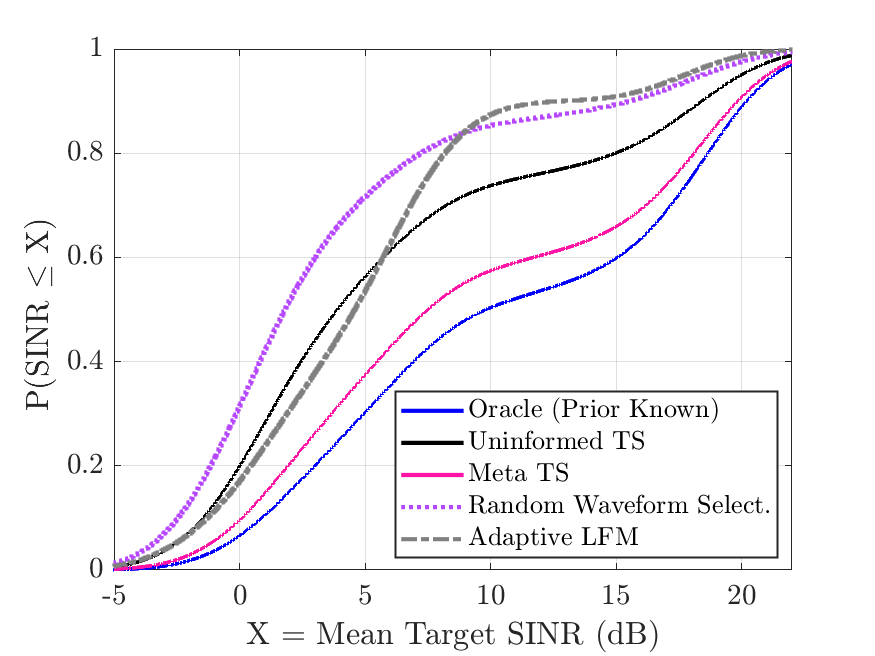}}
    \subfloat[]{\includegraphics[scale=0.55]{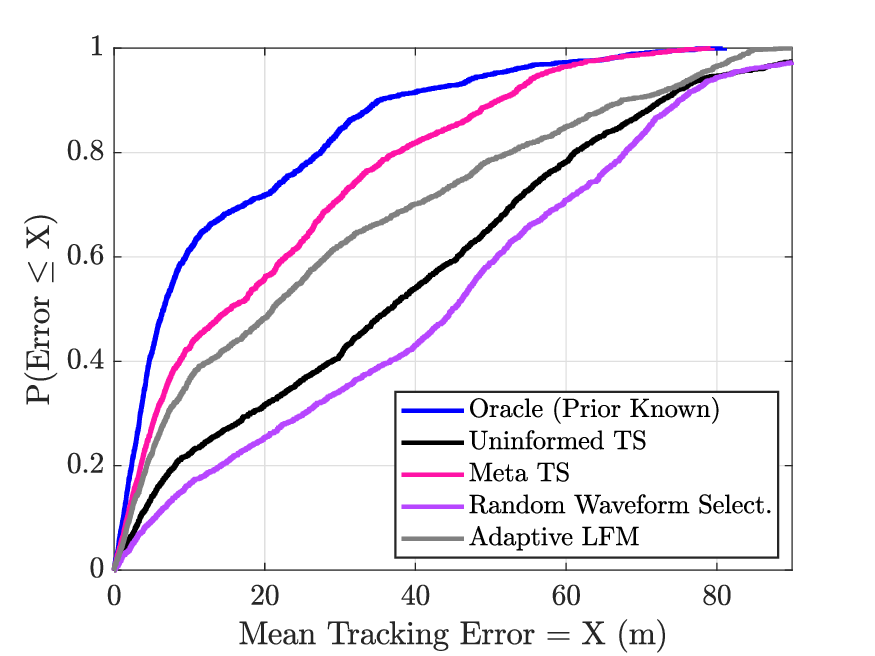}} \\
    \subfloat[]{\includegraphics[scale=0.55]{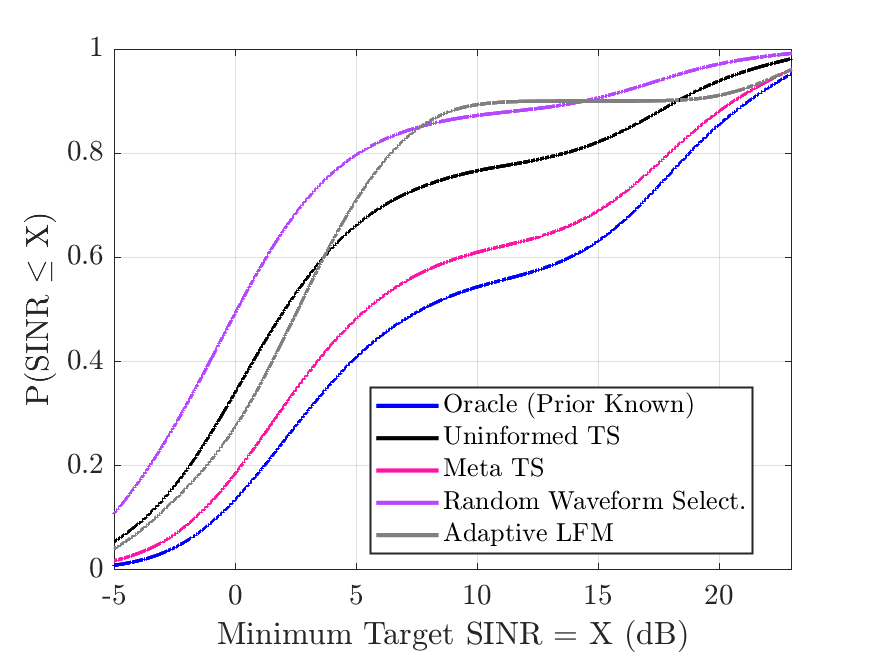}}
    \subfloat[]{\includegraphics[scale=0.55]{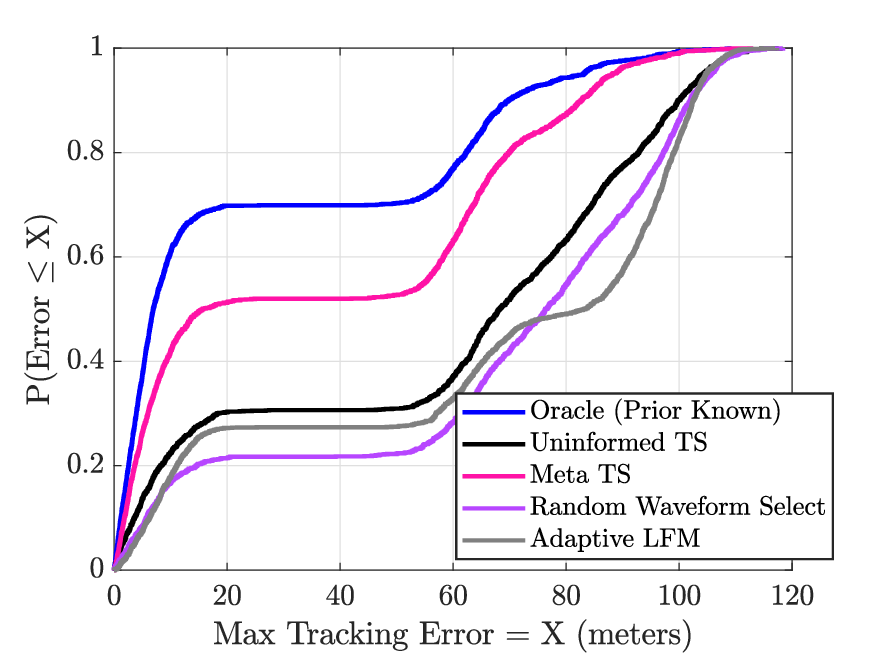}}
	\caption{\textsc{Cumulative outage behavior} of each waveform selection algorithm over a sequence of $N_{\text{track}}=80$ target tracks, each lasting for $n=500$ CPI's. Posterior sampling results in a drastic cumulative performance improvement over random waveform selection. We observe that the most substantial peformance improvements as a result of learning the best prior occur in the regime of $5$dB $ \leq \sinr \leq 15$dB and $60$m $\leq$ tracking error $\leq$ 40m. Learning approaches provide the most substantial benefits over rule-based LFM selection in terms of weakest target performance.}
	\label{fig:cum_outage}
\end{figure*} 

In this section, we examine the performance of the contextual meta-TS waveform selection approach presented in Algorithm \ref{algo:mts}. Meta TS is compared three baseline waveform selection strategies. The first is an `oracle', which performs Thompson Sampling begins with knowledge of the underlying prior distribution. Additionally, we consider a Thompson sampling algorithm which is initialized with an uninformative prior, namely $P(\theta) \sim \mathcal{N}(0_{d},I_{d})$. We further consider a random waveform selection strategy, in which waveforms from the radar's library are transmitted with uniform probability, and a rule-based adaptive strategy which modifies a linear FM waveform depending on the targets' estimated positions. If the average of the estimated target positions is less than $3$Km, the radar transmits an LFM waveform with a bandwidth of $80$MHz. If the average estimated position is $3$Km-$5$Km, the bandwidth of the LFM waveform is $40$MHz. If the average estimated position is greater than $5$Km, the LFM bandwidth is $20$MHz.

We simulate a waveform-agile tracking scenario in which the radar engages in a temporal sequence of target tracks, and adapts its waveform each CPI. Three targets are present in the scene during each tracking instance. Each target tracking instance is defined by a new FSTC parameter $\theta$. We note that during each track, the channel is non-stationary, as the target and clutter impulse responses vary on a CPI-to-CPI basis. However, the general characteristics of the scene remain stationary enough such that $\theta$ is constant. In the simulations, the radar undergoes a sequence of $N_{\text{track}} = 80$ target tracks. Each target track lasts for $n = 500$ CPIs, and each CPI consists of 256 coherent pulses. Assuming a pulse repetition frequency of $10$KHz, each track lasts for approximately 12.8 seconds. The radar has access to a waveform library with $A = 50$ waveforms. The waveform library consists of 10 LFM chirp pulses with a varying chirp rate, 10 Zadoff-Chu phase-coded pulses, 10 Frank coded pulses, 10 Barker coded pulses, and 10 exponential FM pulses with a variable parameter $\alpha$. 

At the beginning of each track, the initial position, velocity, and trajectory of each of the three targets is randomized. The target frequency responses $\kappa(t)$ and the clutter impulse responses $\lambda(t)$ vary from CPI-to-CPI, and have frequency responses drawn from a Gaussian ensemble.

These simulation conditions, along with the loss function and context representation, determine the parameter $\theta$ which the radar wishes to learn. The vector $\theta$ represents a weighting over context features in order to predict the loss associated with each waveform. This underlying structure from track-to-track is what the meta-learning algorithm exploits to effectively bias the algorithm. Over the course of the track, the radar uses its current estimate of the target's position, as well as an estimate of the channel quality, in order to select the best waveform for the next CPI. The loss function used by the radar is 
\begin{equation}
\ell_{k} = \min \left\{\max \left\{ \frac{\sinr_{\texttt{mean}}}{\sinr_{\texttt{target}}}, 0 \right\}, 1\right\},
\end{equation}
where $\sinr_{\texttt{mean}}$ is the SINR, averaged across each of the targets measured from the range-Doppler response. This is computed by performing CFAR detection, clustering the detections using the DBSCAN clustering algorithm \cite{schubert2017dbscan}, and averaging the power over each of the target clusters. The term $\sinr_{\texttt{target}}$ is a target SINR value, which roughly corresponds to the highest SINR the radar can achieve during a given CPI. This loss function is effective since it is bounded from above and below and directly incorporates measurement quality. Further, it is reasonable to assume that a target SINR value will be approximately known for most tracking applications. In these simulations, the target SINR is 22dB.

The set of context vectors $\{\varphi(w_{i},o_{j})\}$ are defined by computing
\begin{align}
	\varphi(w_{i},o_{j}) = \bigg[\E[\ell_{k}(w_{i})|o_{j}], \V[\ell_{k}(w_{i})|o_{j}], \max_{m < k} [\ell_{m}(w_{i})|o_{j}] \bigg],
\end{align}
for each $w_{i} \in \ncalW$ and $o_{j} \in \ncalO$. 

In Figure \ref{fig:outage_time}, we examine the outage behavior of each of the waveform selection strategies over the sequence of $m=80$ tracks. We observe that the Meta TS algorithm quickly approximates the true prior and performs nearly as effectively as the oracle both terms of average SINR, average tracking error, weakest target SINR, and maximum tracking error. Thus it can be maintained that knowledge of the underlying prior drastically improves outage behavior, especially over the limited time horizon of 500 CPIs each track. We also observe that random waveform selection serves the worst performing baseline, with outage events occurring in $30$-$45\%$ of all transmissions. The meta-TS approach generally results in a $30$-$40\%$ reduction outage probability as compared to random waveform selection. We note that the uninformed TS strategy and the rule-based adaptive LFM strategy perform similarly, due to the radar's inability to quickly infer $\theta$ in a short time horizon using an uninformative prior. However, we note that the adaptive LFM approach performs noticeably worse in terms of weakest target SINR and maximum tracking error.

\begin{figure}
	\centering
	\includegraphics[scale=0.55]{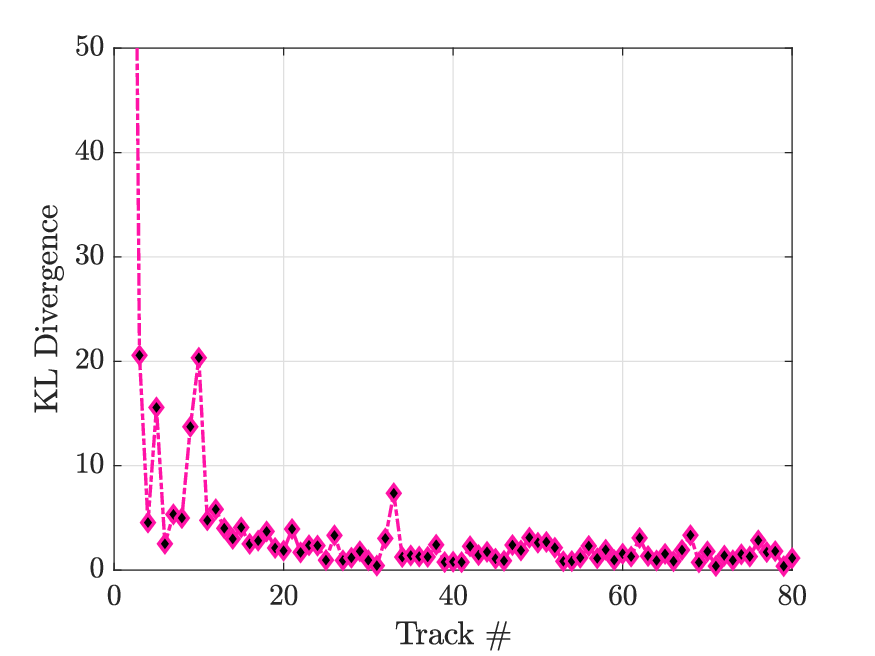}
	\caption{\textsc{Average KL divergence} between the meta-learned prior $Q_{v}$ and the true prior $P^{*}$. After 20 tracks, the meta-learning algorithm has closely approximated the oracle.}
	\label{fig:KL}
\end{figure}

Figure \ref{fig:cum_outage} shows the outage performance, in terms of the previously considered performance statistics, over the entire sequence of 80 tracks. We observe once more that all posterior sampling techniques enjoy a large performance benefit over random waveform selection, which consistently selects waveforms that are poorly matched to the composite target/clutter response of the channel. Additionally, we see that the adaptive LFM strategy performs comparatively poorly in terms of tracking error for the weakest target response.

In Figure \ref{fig:KL} we observe the KL divergence between the meta-posterior $Q_{v}$ and the best prior $P_{\star}$. After just a few tracks, the meta-posterior concentrates toward the true prior in terms of KL divergence, and very closely approximates the oracle after 20 tracks. Thus, we see that the meta-learner is converging toward a distribution which is in line with the true data. We note that the KL divergence does not necessarily need to go to zero in order to achieve acceptable radar performance. The meta-learner's aim is to learn a prior that is effective over each of the possible FSTCs the radar may experience, which is generally feasible even with a small number of examples. Explicitly estimating the true prior, on the other hand, may require far more samples.

\begin{figure}
	\centering
	\includegraphics[scale=0.55]{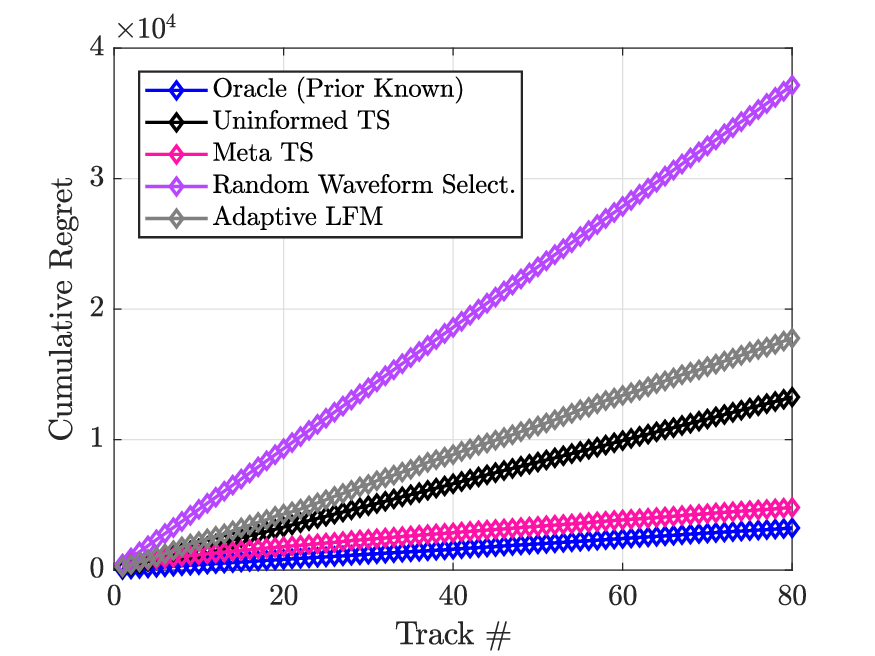}
	\caption{\textsc{Cumulative Regret} of each waveform selection algorithm over the entire sequence of tracks.}
	\label{fig:cumReg}
\end{figure}

\begin{figure*}
	\centering
	\subfloat[]{\includegraphics[scale=0.55]{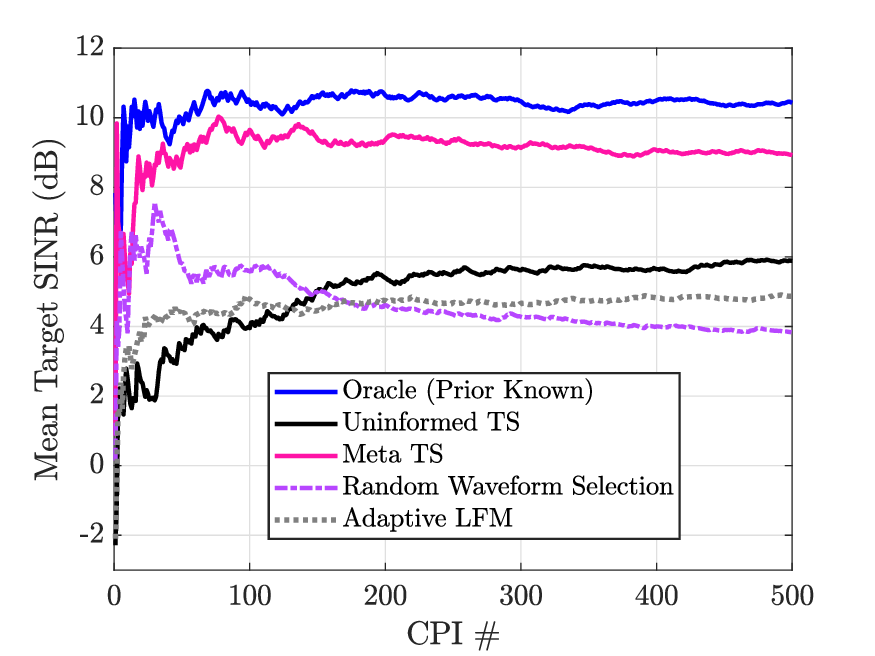}}
	\subfloat[]{\includegraphics[scale=0.55]{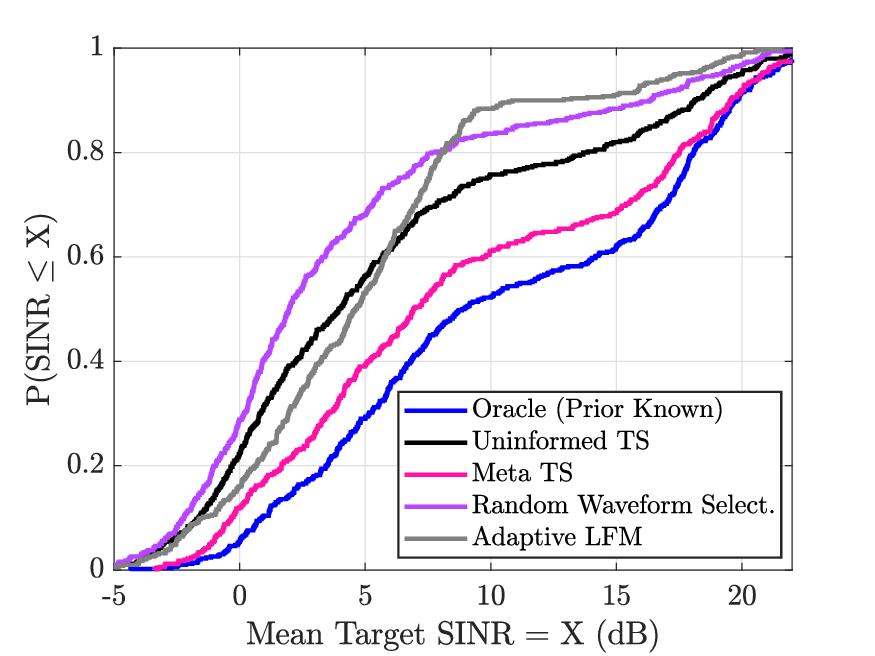}} \\
	\caption{\textsc{Single track behavior} during the $80^{\text{th}}$ and final target track. (a) Cumulative average SINR. (b) Empirical distribution function of mean target SINR.}
	\label{fig:singleTrack}
\end{figure*}

In Figure \ref{fig:cumReg} we observe the cumulative strong regret experienced by each waveform selection algorithm. This performance measure is widely used in the online learning literature and gives a sense of how closely a decision rule follows the best possible strategy under the given problem formulation. We see that when the true prior is known, the radar incurs regret at an approximately constant rate, which is expected given that the true prior is suitable for all FSTCs in the class. Similarly, when an uninformative prior is used, the radar also incurs regret at a nearly constant rate, but much faster given that more exploration is neeeded to learn the dynamics of the new FSTC during each track. When meta-learning is applied, the radar incurs regret at a decreasing rate over time as the true prior is approximated by the meta-TS algorithm. We see that each of the contextual bandit algorithms significantly outperforms the naive random selection strategy, and more narrowly the adaptive LFM strategy, as expected.

In Figure \ref{fig:singleTrack}, behavior over a single track is isolated. We examine performance during the $80^{\text{th}}$ and final target track for each waveform selection algorithm. In particular, we see the cumulative SINR averaged across the three targets for each algorithm on the left side of the figure. We observe that the meta-TS algorithm performs close to the oracle throughout the entire track duration. The uninformed TS algorithm performs significantly worse over the entirety the track, due to the short time horizon, which makes learning without a strong prior difficult. In order to learn effectively with an uninformative prior, the number of waveforms in the catalog would need to be reduced. Thus, we find that meta-learning has greatly improved the learning efficiency in the cases tested. The same conclusion can also be drawn from the empirical distribution function of average target SINR seen on the right side of the figure, from which it is observed that meta-learning provides substantial gains in the $5$dB $\leq$ SINR $\leq$ $15$dB regime once again.

\section{Conclusion}
\label{se:concl}
This work has presented an online meta-learning approach for waveform-agile radar tracking. This approach allows the radar to learn transmission policies from a limited number of interactions by exploiting inherent similarities in the physical scene from track-to-track. We have defined the waveform selection problem as a sequence of Bayesian learning tasks, where each learning instance corresponds to a new target track. Each track, the radar sequentially chooses waveforms based on an estimate the posterior distribution of a finite-state target channel coming from a fixed class. We have observed that misspecification of the prior distribution can cause significant performance degradation, motivating the study of algorithms which can automatically learn a prior which works well across the entire class of channels. 

In essence, the radar learns about the underlying structure of the physical environment that carries over from one track to the next. This information is used to \emph{bias} the learning process so that less exploration is required to develop a transmission policy. The implications for target tracking are as follows:

\begin{itemize}
	\item Less exploration at the beginning of each track drastically reduces the occurrence of `outage' events that may result in a lost track.
	\item An initial bias allows for the use of larger waveform catalogs, which provide more diversity.
	\item High-level bias information could be communicated and used to inform other radars or processes in a distributed network.
\end{itemize}

Thus, the problem of bias learning has direct practical applications, especially in terms of reducing worst-case performance when the scene has fundamental characteristics which remain relatively stationary over time. However, we note that meta-learning may not be appropriate for all radar tracking applications, and robust systems are expected to use an array of rule-based and learning approaches to reduce worst-case performance in unexpected conditions.

Herein, we have shown that the problem can be effectively formulated in terms of Bayesian learning, for which a computationally feasible contextual bandit algorithm is developed. This approach is satisfying as the online waveform selection naturally fits well within the contextual bandit framework. However, the Bayesian interpretation of meta-learning is not exclusive, and the problem can also be approached from a functional viewpoint \cite{Baxter1998}. Future research comparing the Bayesian and functional interpretations would be of value. Additionally, an investigation into the sharing of bias information in a distributed network is an avenue for future work. Knowledge of spatial correlation structure could significantly speed up bias learning.

\appendices
\section{Computation of the Posterior $P(\theta|\ncalF_{k-1})$}
\label{se:post}
Let the prior distribution be $\mathcal{N}(\hat{\theta}_{k},B_{k}^{-1})$, where $\hat{\theta}_{k} \in \nbbR^{d}$ and $B_{k} \in \nbbR^{d \times d}$ are the mean vector and covariance matrix of a multivariate normal distribution. Further, let $\ell_{k}(w_{i})$ be the loss associated with transmitting waveform $w_{i}$ at step $k$. Then we may compute the posterior distribution as

\begin{align}
	\operatorname{P}(\theta \mid \ell_{k}(w_{i})) &\propto  \operatorname{P}\left(\ell_{k}(w_{i}) \mid \theta \right) \operatorname{P}(\theta) \\
	&\propto \exp\{-\frac{1}{2}((\ell_{k}(w_{i})-\theta^T \varphi_{i,k})^2 \\
	& \quad+(\theta-\hat{\theta}_{k})^T B_{k}(\theta-\hat{\theta}_{k})\} \nonumber \\
	&\propto \exp \{-\frac{1}{2} (\ell_{k}(w_{i})^2+\theta^T \varphi_{i,k} \varphi_{i,k}^T \theta \\
	& \quad+\theta^T B_{k} \theta-2 \theta^T \varphi_{i,k} \ell_{k}(w_{i})-2 \theta^T B_{k} \hat{\theta}_{k})\} \nonumber \\
	&\propto  \exp \{-\frac{1}{2}(\theta^T B_{k+1} \theta-2 \theta^T B_{k+1} \hat{\theta}_{k+1})\} \\
	&\propto \exp \{-\frac{1}{2}(\theta-\hat{\theta}_{k+1})^T B_{k+1}(\theta-\hat{\theta}_{k+1})\} \\
	&\propto  \mathcal{N}(\hat{\theta}_{k+1}, B_{k+1}^{-1}) .
\end{align}

\bibliographystyle{IEEEtran}
\bibliography{sensingBib}{}

\end{document}